# COMPRESSION WAVES AND PHASE PLOTS: SIMULATIONS

## D. Orlikowski[1] and R. Minich

[1]L-45, 7000 East Ave., Livermore CA 94551

**Abstract.** Compression wave analysis started nearly 50 years ago with Fowles.[1] Coperthwaite and Williams [2] gave a method that helps identify simple and steady waves. We have been developing a method that gives describes the non-isentropic character of compression waves, in general.[3] One result of that work is a simple analysis tool. Our method helps clearly identify when a compression wave is a simple wave, a steady wave (shock), and when the compression wave is in transition. This affects the analysis of compression wave experiments and the resulting extraction of the high-pressure equation of state.

## INTRODUCTION

To gain and understand high-pressure equation of state (EOS), ramp compression is applied to one side of a material. Stress or particle velocity measurements are made at different Lagrangian points either in the sample or at an interface. With these measurements, methods exist to extract EOS information since Fowles et al. and Cowperthwaite and Williams.[1,2] Recently, ramp compression experiments are becoming common [4,5] and exceeding 10s of Mbar in pressure.[6] Because the compression thermodynamic trajectories are unknown, but are not shocks, the question of how much entropy is being generated, if at all, arises. Similarly, is it possible currently to identify when standard analysis cannot be applied.

Current, analysis techniques overlook the fundamental assumptions on which those procedures are based. To extract the stress-density ($\sigma$-$\rho$) relationship from a pair of particle velocities $u(h,t)$ over time t at two Lagrangian points $h_1$ and $h_2$, the Lagrangian sound speed is determined with $C_L(u) = \delta h/\delta t$ at constant u. Then, a $\sigma$-$\rho$ relationship can extracted via $d\sigma = \rho_0 C_L(u)\, du$ and $d\rho = du\rho_0/C_L(u)$, which derives from the Riemann integral at constant entropy. From the characteristics point of view, it is the same as characteristics of constant u or $\rho$ are linear in the h-t plane. However, to reach these relations from the conservation of mass and momentum equations, assumptions are made: 1) *There is no heat conduction in the system;* 2) No *entropy is generated, i.e. no dislocations, no viscosity, etc.*

We describe a method that does not make assumptions and offers insight into the compression wave character, with which analysis is better defined. The method identifies steady and/or simple waves over $\Delta h$. A compression wave can be steady, and yet be isentropic or non-isentropic, e.g. a shock. If there is no phasing difference between sound speeds $C_\sigma$ at constant stress and $C_u$ at constant u, then a simple wave exists. We derive a method and then use this method to analyze hydrodynamic simulations of an isentropic compression and a compression wave that evolves through the material. The method helps to clearly identify regions in the u-histories where EOS information may be extracted.

## METHODOLOGY

In two sections we present an outline of the relationship between Lagrangian sound speed and particle velocity (*in situ*) and also a description of the hydrodynamic simulations.

### Derivation

To extract the stress-density ($\sigma$-$\rho$) relationship along an *isentropic* thermodynamic path, the Lagrangian sound speed is fundamental, because $d\sigma = \rho_0 C_L(u)\, du$ and $d\rho = du\rho_0/C_L(u)$, which derives from the Riemann integral at constant entropy. *If entropy is generated, then these relationships cannot be used to extract EOS information, other methods must be used.*[3] To obtain $C_L(u)$, the relationship $\delta h = C_L(u)\, \delta t$ is commonly used, where $\delta t = t_+ - t$ between two u-histories, e.g. across a step on the back surface of an experimental target. In Fig. 1 we show a typical example of particle velocity history with $\delta t$ at constant u. All $u(t)$ in this paper are *in situ*; we want to avoid further complicating interaction with window or free surface interfaces, that are necessary in experiments. To clearly identify the linear and non-linear contributions, we write $C_L(u)$ as $C_L(u) = c_0 + c_1 u + F(u,t)$, where $c_0$ and $c_1$ are coefficients and $F(u,t)$ is a function accounting for the non-linear contribution. Therefore, taking the derivative of $\delta t$ with respect to $du$ and defining $\Gamma^{-1} = (du/dt)^{-1}$, we then write

$$\Gamma_+^{-1} - \Gamma^{-1} = (h_+ - h)\frac{d}{du}C_L^{-1}(u). \quad (1)$$

The non-linear contribution can be identified as

$$\Gamma_+^{-1} - \Gamma^{-1} = (h_+ - h)\frac{d}{du}(C_u^*)^{-1} + G(u) \quad (2)$$

$$G(u) = \frac{d}{du}\int_t^{t_+} dt\, C_L \frac{F(u,t)}{C_u^*}, \quad (3)$$

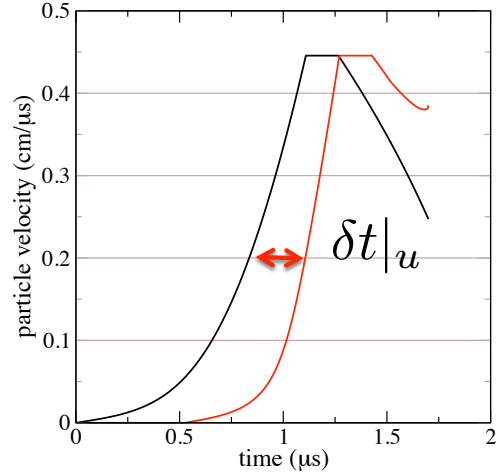

**Figure 1.** Typical *in situ* particle velocity u-history is show; the time difference at two different thicknesses within the bulk material. In an experiment this would be at interfaces; those interfacial u need to be backward propagated to the *in situ* u.

where $C_u^* = c_0 + c_1 u$ is used. With this representation we are able to quantify the non-linear contribution to the $C_L(u)$. Additionally, Eqn. 1 identifies clearly the formation of shocks, discussed below.

### Simulation

To simulate a compression wave, we used the LLNL hydrodynamic code CALE.[7] We applied a smooth compression wave to a one dimensional material, which was semi-infinite so as to remove interface effects. The *in situ* u(t) are extracted at particular Lagrangian points at every 0.04 cm from 0.04 to 0.64 cm with total length being 1.28 cm.

The material EOS for vanadium is used from the standard database LLNL EOS 9231, developed at LLNL.[8,9] Vanadium was chosen based on fine resolution of the numerical table, the lack of any phase transition in this particular LEOS, and method of development of the EOS. To maintain simplicity in the wave evolution, we did not use a strength model.

To simulate behavior of entropy production within a compression wave, i.e. the entropy necessary to evolve wave into a shock wave, we used Von Neumann artificial viscosity [10] to our advantage, which accounts for the necessary entropy discretely across a zone in the mesh. For the artificial viscosity model we used $C_Q = 1.9$ and $C_L = 0.75$, which is higher than normal, but this is to have faster steepening of the wave only. Initially, we had a very fine mesh of 15,000 zones/cm, which effectively turns off the artificial viscosity giving isentropic compression. This vetted our formulations. Then, we coarsened the mesh to 150 zones/cm obtaining strong heating as function of distance through the material, as discussed below.

## RESULTS AND DISCUSSION

We discuss the particle velocity analysis using our formulation for two cases: the perfect isentropic compression wave and the compression wave that evolves into a shock over distance with evolving temperature.

We first verify our methodology of analyzing the simulated u(t). We used 16 u-histories and took combinations yielding 136 pairs. Rewriting Eqn. 1 as $\Delta\Gamma^{-1}/\Delta h = dC^{-1}/du$, we can collapse all combinations onto one curve (See Fig. 2). This verifies that the simulated u-histories do reproduce the theoretical, isentropic $C_L(u)$ that is extracted from the vanadium EOS. Because the waves are scale invariant of $\Delta h$, then Eqn. 1 can be integrated to obtain EOS information as previously mentioned. Moreover, this $\Delta h$ scale invariance means that $\partial C_L(u)/\partial t = 0$, implying that stress is only a function of $u$ – a simple wave.

Additionally, we compare the linear representation of the $C_L(u)$ to those results to understand if the non-linear contribution can be identified. The non-linear contribution results are not shown because of space requirements; however, by subtracting $dC_u^{-1}/du$ from $\Delta\Gamma^{-1}/\Delta h$, the function $G(u)/\Delta h$ can be determined.

With the methodology verified, we wanted to simulate a wave evolving into a shock over a distance within the material. As described in the methodology section, we used the Von Neumann artificial viscosity with a coarsened mesh. We then sampled the material very earlier, mid-way and at end in its evolution, i.e. respectively, nearly isentropic, mid-trajectory in thermodynamic space, and full shock wave. These sampled stages correspond to different thermodynamic trajectories through the phase diagram depicted in Fig. 3d. The resulting analysis shown in Fig. 3 indicates the difficulty in obtaining a reliable $dC_L^{-1}/du$ from the simulated data. In Fig. 3a we were able to extract one combination pair from the u-history that is close to the theoretical $dC_L^{-1}/du$. But this is only by chance. Because the self-similar attribute of the compression wave is broken immediately as the wave evolves over some distance, the basis of our derived relationship Eqn. 1 is destroyed. That is $\partial C_L(u)/\partial t \neq 0$, i.e. the system is non isentropic. Additionally, the relations $d\sigma = \rho_0 C_L(u)\,du$ and $d\rho = du/\rho_0 C_L(u)$ cannot be used for EOS, because the wave is no longer isentropic.

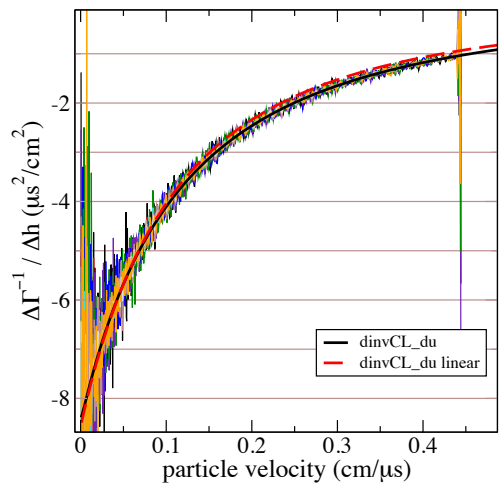

**Figure 2.** Analyzed u-history pairs (136 in total) via our method give the expected theoretical relation of $dC_L^{-1}(u)/du$ (black solid line). The linear approximation $C^*_u$ (red dashed line) demonstrates the sensitivity required. At u = 0.4 cm/ms, the pressure is approximately 3 Mbar.

However, the interesting aspect of this formulation (Eqn. 1) is the clear identification of the compression wave evolving into a shock. Because it identifies as a function of u non-isentropic behavior, it also identifies when the u(h,t)-characteristics are non-linear. As the wave steepens into a shock the value of $\Delta\Gamma^{-1}/\Delta h$ for each u approaches the zero. In the limit of a shock, $\Delta\Gamma^{-1}$ becomes zero because the shock wave stops evolving (shown in Fig. 3c), i.e. a steady wave.

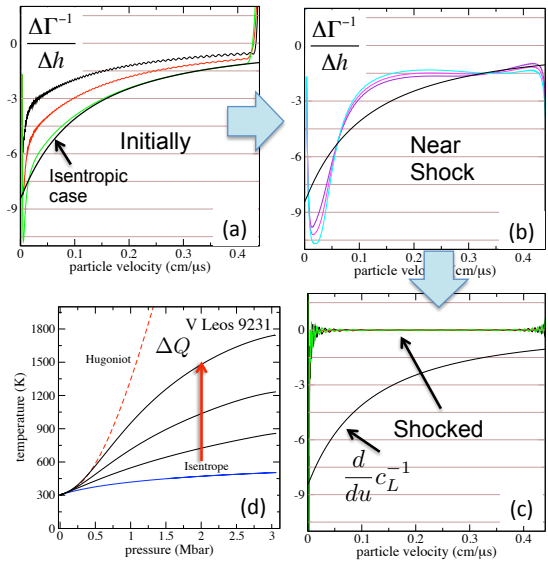

**Figure 3.** For the case of an evolving wave into a shock, we analyzed different locations in the wave evolving into a shock.

## CONCLUSIONS

We present a method that clearly identifies regions in compression waves, where the wave is evolving. We demonstrated the method for two simulated cases: an isentropic compression wave and a non-isentropic, evolving compression wave. We simplified the system to minimize confusion with other models, i.e. no strength models. The results indicate that common analysis methods to obtain EOS are adequate when the compression wave is isentropic. However, for non-isentropic waves those tools are no longer applicable. We provide a method that clarifies those cases. Moreover, if Δh is small enough, then the sampled waves may appear steady. This would give conflicting σ-ρ relationship, because the wave is non-isentropic breaking the mathematical requirement of the isentropic equations. Therefore, more than 3 samples of Δh over longer h to clearly identify if $\partial C_L(u)/\partial t = 0$ in the system before any analysis can be done. This affects EOS analysis and backward integration techniques [11] resulting in an over-estimate of *in situ* stress. In this case, three families of characteristics are required for the Lagrangian representation in contrast to the two families often used in backward wave integration techniques. [12]

The underlying assumptions of no heat conduction and of no relaxation processes are too restrictive. These assumptions mean no velocity gradients (no viscosity); dynamic material response is mostly plastic deformation, phase transitions, constrained fracture, grain sliding, etc. To this larger issue of analyzing non-steady compression waves, we are working on method to extract the necessary change in entropy.[3] This will enable a clearer understand of the wave evolution and the resulting extracted EOS information.

## ACKNOWLEDGEMENTS

This work performed under the auspices of the U.S. Department of Energy by Lawrence Livermore National Laboratory under Contract DE-AC52-07NA27344.